\documentstyle[12pt]{article}
\renewcommand\caption{}

\title{On the Effective Potential for Local Composite Operators
}

\author{{\sl E. V. Gorbar} \\
{\sl{Instituto de Fisica Teorica, 01405-900 Sao Paulo, Brazil} \dag}
}

\date{}

\begin{document}

\maketitle

\vfill

\begin{abstract}

We show that the effective potential for local composite operators is a useful object in studing
dynamical symmetry breaking by calculating
the effective potential for the local composite operators $\bar{\psi}
\psi$ and $\phi^2$ in the Gross--Neveu (GN) and O(N) models, respectively.
Since the effective potential for local composite operators can be calculated by using the
Cornwall--Jackiw--Tomboulis (CJT) ef\-fective po\-ten\-tial in theory with additional bare mass terms,
we show that divergences in the effective potential for local composite operators are the
same as in the CJT effective potential. We compare the results obtained with the results give by
the auxiliary field method.
 
\end{abstract}
\dag On leave of absence from Bogolyubov Institute for Theoretical Physics, 252143,
Kiev, Ukraine

\vspace{2mm}

\vfill
\eject

\newpage
\centerline{I.  Introduction.}

\vspace{1cm}

Dynamical symmetry breaking is a difficult problem which 
requires the use of nonperturbative methods of investigation.
The effective potential for local composite operators seems a quite natural generalization of the usual
effective potential for elementary scalar fields \cite{Sym, Colema}. However, it is not the way in which
people usually study dynamical symmetry breaking. The reason is that as it was shown \cite{Banks}
(see also \cite{Ell})
that unlike the effective potential for elementary scalar fields the effective potential for local composite
operators
is plagued with ultraviolet divergences outside stationary points. The point is that the effective potential
for elementary fields is finite if the usual renormalization of fields, coupling constants, and masses is performed.
This is not true for the effective potential for
local composite operators which is divergent even if the usual renormalization procedure is performed.
Therefore, the method of the
effective potential for local composite operators was practically
abandoned 
for long time and the method of bilocal
composite operators [5-7] (for a review see book \cite{mirbook}) was widely used. However,
recently, the usefulness of the effective action
for local composite operators has been advocated \cite{Mir}. In
 particular, it was applied to the gauged
Nambu--Jona-Lasinio (NJL) model and it was shown that the method works even in the cases where
the auxiliary field method fails. After that the method was successfully applied to the description
of the conformal phase transition in gauge theories \cite{MYa} and to the description of the
nonperturbative dynamics in $QED_{3}$ \cite{GMSh}.

In this paper we continue the study of the effective potential for local composite
operators by
calculating the effective potential for the local composite operator $\bar{\psi}\psi$ and $\phi^2$
in the GN and O(N) models, respectively. Since the effective potential for local composite
operators can be calculated by using the
Cornwall--Jackiw--Tomboulis (CJT) effective potential \cite{CJT} for bilocal operators taken at its extrema
in the theory with additional
source terms for the corresponding local composite operators, we show that actually
divergences in the effective potential
for local composite operators are the same as in the CJT effective potential taken at its extrema.
Therefore, we argue that the negative attitude with respect to the effective potential for local
composite operators due to the problem of ultraviolet divergences mentioned above have to be reconsidered.

As well known
the auxiliary field method is widely used in investigation of some physical models (like, e.g.,
the GN \cite{Gross} and O(N) models \cite{Cole};
for a recent review of these models see,
e.g., \cite{Zinn}). The introduction of auxiliary field
allows one to rewrite the Lagrangian in a quadratic form of the
initial fields.
Then, by integrating over the initial fields, we obtain the
effective action for auxiliary field. Note that the
GN model is exactly solvable (its exact
S-matrix was found in \cite{Zam}.)
We compare the method of the effective potential for local composite operator with the auxiliary field method.
In agreement with the paper \cite{Mir}, we show that the method
of the effective potential for composite operators successfully works even in the cases where the
auxiliary field method fails.

In the GN model it was shown long ago (see
\cite{Gross, Hay}) at $g > 0$
the effective potential for the local composite operator
$\bar{\psi}\psi$ and
the effective potential for auxiliary field describe essentially the same physics and coincide in the continuum limit
(cut-off $\Lambda^2 \to \infty$). However, we show that at $g < 0$
the effective potentials have a different behavior. The effective potential for the local composite operator $\bar{\psi}
\psi$ is a monotonously increasing function of $<0|\bar{\psi}{\psi}|0>$, meanwhile, the effective potential for auxiliary
field is a monotonously decreasing function of $\sigma$. In this case the
auxiliary field does not describe the dynamics of any physical state of the model. However, it can be useful in getting
S-matrix because its propagator describes fermion-antifermion scattering.
We briefly comment on the physical meaning and interpretation of the effective potentials. Since $<0|\phi_c^2|0>$ is not
an order parameter in the O(N) model, the effective potential for the
local composite operator $\phi^2$ is less useful than 
the effective potential for $\bar{\psi}\psi$ in the GN model.
\vspace{1cm}

\centerline{II.  The Gross--Neveu model.}

\vspace{1cm}

Let us calculate the effective potential
for the local composite operator $\bar{\psi}\psi$ in the GN model.
The Lagrangian of the GN model \cite{Gross} reads
\begin{equation}
L = i\sum_{k=1}^{N}\bar{\psi_k} \gamma^\mu \partial_\mu
\psi_k +\frac{g}{4N} (\sum_{k=1}^{N}\bar{\psi_k}\psi_k)^2,
\end{equation}
where N is the number of flavors and the dimension of spacetime
is 1 + 1.
Lagrangian (1) is
invariant with respect to the discrete chiral symmetry ($\psi \rightarrow
\gamma_5 \psi$). The calculation of the effective potential proceeds as follows.
By adding the source $J\bar{\psi}\psi$ and
integrating over the fermion fields, we obtain the
generating functional
\begin{equation}
e^{iW(J)} = \int D\bar{\psi}D\psi e^{i\int(L + J\bar{\psi}\psi)d^2x}.
\end{equation}
The $\sigma_c$ field (a classical field) is
\begin{equation}
\sigma_c = <0|\bar{\psi}\psi|0> = \frac{\delta W(J)}{\delta J}.
\end{equation}
The effective action for $\sigma_c$ is given by the Legendre
transformation of $W(J)$
\begin{equation}
\Gamma(\sigma_c) = W(J) - \int J \sigma_c d^2x.
\end{equation}
To obtain the effective potential, it suffices to set J = const.
Then the relation $\Gamma (\sigma_c) = - \int d^2x V(\sigma_c)$ gives us the effective potential
$V(\sigma_c)$ .
We calculate $V(\sigma_c)$
by integrating the obvious equality $\frac{d V(\sigma_c)}{d\sigma_c} = J(\sigma_c)$.
To get
$V(\sigma_c)$, we should express $J$ through $\sigma_c$. In the case under consideration it is
very difficult to perform. Therefore, by following paper \cite{Mir}, where
the effective potential in the gauged NJL
model was first calculated, we use
the fermion mass as a variable in the effective potential instead of the variable $\sigma_c$
\begin{equation}
\sigma_c = <0|\bar{\psi}\psi|0> = -N\frac{m}{2\pi}\mbox{ln}(\frac{\Lambda^2}{m^2}
+ 1).
\end{equation}
Thus, we have
\begin{equation}
V(\sigma_c) = \int J(\sigma_c)d\sigma_c = \int J(m) \frac{d\sigma_c}{dm} dm.
\end{equation}
The most difficult part is to calculate the generating function $W(J)$
or what we need is our case (see Eq. (3)) a relation between $J$ and $\sigma_c$ ($\sigma_c$ is equal to
$\frac{\delta W}{\delta J}$, therefore, all the same we need to know $W(J)$).
Let us show that one can actually calculate $W(J)$
by using the CJT effective potential for bilocal operators. Indeed, what is the physical meaning of $W(J$)?
At constant $J$ it is the vacuum energy density integrated over spacetime of the model with additional bare mass
term $J \bar{\psi}\psi$. It is easy to see how it can be calculated
from the CJT effective potential.
Indeed, it is nothing else but the value of the CJT effective potential (in the system with that additional bare mass
term) taken at its extremum. As well known the CJT effective potential at its extremum gives
the Schwinger--Dyson (SD) equation. This
gives us the required relation between $J$ and $m$
\begin{equation}
J = \frac{gm}{4\pi}\mbox{ln}(\frac{\Lambda^2}{m^2} + 1) - m,
\end{equation}
which is the SD equation at first order in $g$
in model (1) with the bare mass term $J \bar{\psi}\psi$.
Thus, from (5)-(7) we find $V(m)$
\begin{equation}
V(m) = \frac{N}{4\pi} \int (1 + \frac{g}{4\pi}\mbox{ln}(\frac{\Lambda^2}
{m^2} + 1))(\mbox{ln}(\frac{\Lambda^2}{m^2} + 1) - \frac{2}{1 + \frac{m^2}
{\Lambda^2}})dm^2.
\end{equation}
Finally, by integrating in (8) we obtain the effective potential
\begin{equation}
V(m^2) = \frac{N m^2}{4\pi}(-\frac{g}{4\pi}\mbox{ln}^2
(\frac{\Lambda^2}{m^2} +
1) + \mbox{ln}(\frac{\Lambda^2}{m^2} + 1) - \frac{\Lambda^2}{m^2} \mbox{ln}(\frac{m^2}{\Lambda^2} + 1)).
\end{equation}

Before analysing the effective potential (9) we outline the calculation of the effective potential
for auxiliary field in order to compare in what follows the results given by two effective potentials.
By using the Hubbard--Stratonovich auxiliary field method \cite{Hub, Stra}, we
represent Lagrangian (1) in the equivalent form
\begin{equation}
L= \sum_{k=1}^{N}(i\bar{\psi_k} \gamma^\mu \partial_\mu\psi_k + \sigma\bar{\psi_k}\psi_k)
- N\frac{\sigma^2}{g}
\end{equation}
(according to Lagrangian (10), the equation of motion for the
auxiliary field is $\sigma = \frac{g\bar{\psi}\psi}{2N}$ and substituting it in (10)
we obtain Lagrangian (1)).

By integrating over the fermion fields and setting $ \sigma = $ const, we obtain
the effective potential for the auxiliary field $\sigma$
\begin{equation}
\frac{V_{af}(\sigma)}{N} = \frac{\sigma^2}{g} - \int \frac{d^2p}{(2\pi)^2} \mbox{ln} (1 +
\frac{\sigma^2}{p^2})= \nonumber \\
\frac{\sigma^2}{g} - \frac{1}{4\pi}\left[\Lambda^2\mbox{ln}(\frac
{\sigma^2}{\Lambda^2} + 1) + \sigma^2\mbox{ln}(\frac{\Lambda^2}{\sigma^2} + 1)\right].
\end{equation}

It is easy to see that  there is a nontrivial minimum ($\sigma^2 = \Lambda^2 e^{-\frac{4\pi}{g}}$)
of $V_{af}(\sigma)$ at $g > 0$.
On the other hand, at $g < 0$ $V_{af}(\sigma)$ is a monotonously decreasing function and
does not have a minimum. The absence of a stable vacuum means that
in this case the field $\sigma$ is not a "good" variable and does not describe nontrivial dynamics of any
physical states
of the system. This is in contrast to the the case $g > 0$
where $V_{af}(\sigma)$ has a nontrivial minimum and where the field $\sigma$
describes the dynamics of a bound state of the model in the fermion-antifermion channel
(since $g > 0$, we have an attraction in the fermion-antifermion channel that in two-dimensional
space-time immediately leads to the formation of a bound state).
Indeed, on the equation of motion $\sigma = \frac{g\bar{\psi}\psi}{2N}$ and,
as well known \cite{Haag}, any local composite operator with right quantum numbers can be chosen
as an interpolating field for the corresponding
bound state.  On the other hand, at $g < 0$ we have
repulsion in the fermion-antifermion channel and the corresponding bound state is absent. Therefore,
in this case the field $\sigma$ does not describe the dynamics of any physical states of the system and its introduction
is only a technical trick.

The difference between the cases of $g > 0$ and
$g < 0$ is even more evident in Euclidean space. Since the integral $\int_{-\infty}^{+\infty}
e^{-\frac{i\sigma^2N}{g}} d\sigma$ exists both for
positive and negative $g$, in Minkowski space potential (11) is valid for any sign of $g$. However, this is not true
in Euclidean space. For $g > 0$ we can introduce the auxiliary field $\sigma$ in the usual way because the integral
$\int_{-\infty}^{+\infty}e^{-\frac{\sigma^2N}{g}} d\sigma$ exists for positive $g$. However, it is not immediately
clear how one can introduce auxiliary field at $g < 0$ because the integral
$\int_{-\infty}^{+\infty}e^{-\frac{\sigma^2N}{g}} d\sigma$  is meaningless for
$g < 0$. Does it means that at $g < 0$ it is not possible to introduce an auxiliary field in Euclidean space
and thus the formulations of the model in Minkowski and Euclidean spaces are not equivalent? No, it does not.
We can introduce an auxiliary field at $g < 0$ by integrating over $\sigma$ along the imaginary axis (see \cite{Zinn},
p.33). Indeed, $\int_{-i\infty}^{+i\infty} d\sigma e^{-\frac{\sigma^2N}{g} + b\sigma} = i\sqrt{\frac{\pi|g|}{N}}
e^{\frac{b^2g}{4N}}$ ($g < 0$) and, by setting $b = \bar{\psi}\psi$, we rewrite the Lagrangian in a quadratic form in the
fermion fields. However, although we managed to introduce the auxiliary field, in this case the $\sigma$ field
is not a usual field. Indeed,
we integrate along the imaginary axis, however,
the classical equation of motion gives real $\sigma = \frac{g\bar{\psi}{\psi}}{2N}$.
This clearly shows that at $g < 0$ the $\sigma$ field is not a "good" variable
and does not describe the dynamics of any physical states of the system, however, it is useful in getting S-matrix
because the propagator of $\sigma$ describes fermion-antifermion scattering.

Let us study the behavior of the effective
potential for the local composite operator (9) as a function of $m^2$ for small $m^2$ ($m^2 \ll \Lambda^2$).
It is easy to show that at $g > 0$ there is a non-trivial minimum of $V$ at
$m^2 = \Lambda^2 e^{-\frac{4\pi}{g}}$. It coincides with the minimum given by the effective potential
for auxiliary field $V_{af}$ (11) and with the corresponding result in \cite{Gross}. This is in accord
with the conclusion of
\cite{Gross}, where it was shown
that in the continuum limit $\Lambda^2 \to \infty$ the effective potential
for the composite field coincides with the effective potential for the auxiliary field at $g > 0$ when the GN
model is asymptotically free. However, as Gross and Neveu noted, in non-asymptotically free
theories the two effective potentials do not
coincide in general. This is true for the GN model at $g < 0$ (this is directly seen
from comparison of the effective potentials (9) and (11)), where the model is not asymptotically free.
The effective potential for the composite operator (9) gives a reasonable result because it is a growing
function of $m^2$ at least for small $m^2$ (a natural result in the case of repulsion). (For $m^2 \gg \Lambda^2$
the effective potential (9) tends to infinity for any value of $g$. This is connected with well-known fact that
the energy density for fermions is unbounded from below as $m^2 \to \infty$ even in free theory
(see, e.g., \cite{Pes} for the case of the CJT effective potential)).
On the other hand, the effective potential for auxiliary field
$V_{af}(\sigma)$ (11) is a monotonously decreasing function of $\sigma$. This is not a behavior
which one would expect in the case of repulsion in the corresponding channel.
As we noted above this means that at $g < 0$ the introduction of auxiliary field is a
technical trick and the field $\sigma$ in this case does not describe the dynamics of any physical states of the system.
However, as we noted above it can be
used for getting S-matrix because the $\sigma$ propagator describes fermion-antifermion scattering. Consequently, we can
consider the effective actions for local composite operators and auxiliary fields as complementary.

Let us comment on the physical meaning of two effective potentials.
The effective potential for auxiliary field has, in general, a physical meaning only
at stationary points (at
$g > 0$ the physical meaning of the field $\sigma$ is wider because in this case it describes
the fermion-antifermion bound state). On
the other hand, the effective potential for the local composite operator (9) is
physically meaningful also outside stationary
points because the composite operator $\bar{\psi}(x)\psi(x)$ is local. Indeed, it is well known \cite{Banks, HP} that
potentials for nonlocal composite operators have a simple interpretation only at stationary points where they are
equal to vacuum energy densities. In general, at other points they correspond to having nonlocal sources turned on
(for example, $\bar{\psi}(x)J(x,y)\psi(y)$ in the case of the CJT effective potential) and if these sources are nonlocal
in time their interpretation as energy densities breaks down and they are, in general, unbounded from below (this is
the case for the CJT potential [20-24]). However, if we consider the
effective potential for local operators or at least for sources which do not dependent on time, then the corresponding
potential is physically meaningful outside stationary points \cite{Banks}.

As we noted in Introduction the method of the effective action for local composite operators was not widely used
because of the presence of ultraviolet divergences. Instead,
the CJT effective action for bilocal operators was
used. However, as we showed above one can actually calculate the effective potential for local composite operators
by using the CJT effective potential taken at its extrema in the system with the corresponding
additional source terms. This shows that in fact divergences in the effective potential
for local operators are the same as in the CJT potential taken at its extrema.
One of reasons why it was not noted earlier is connected with the choice of variables used.
Usually the mass function $B(p^2)$ is
used as a
variable in the CJT effective potential and $<0|\bar{\psi}\psi|0>$ as a variable in the effective potential for
local composite operators, which may diverge even if $B(p^2)$ is finite. Thus, we
confirm the results of \cite{Mir} that the effective potential for local composite
operators is a very useful object in studing dynamical symmetry breaking and it works even in the cases where the
auxiliary field method fails. In fact, the effective potential for local composite operators has an advantage over the
CJT effective potential for bilocal operators. The CJT effective potential gives the energy density of the system only at
stationary points
because as we noted above outside stationary points nonlocal sources are turned on and the criterion of energy
stability cannot be used. On the other hand, the effective potential for local composite operators gives the energy
density of system with the corresponding constrainted vacuum expectation values of composite operators.

\vspace{1cm}

\centerline{III. O(N)-model.}

\vspace{1cm}

In this section by using as an example the O(N) model we show that the method for the calculation
of the effective potential for local composite
operators can be easily extended to the case of scalar fields.
The Lagrangian of the O(N) model \cite{Cole} reads
\begin{equation}
L = \sum_{k=1}^{N}(\frac{1}{2}\partial_{\mu}\phi^k\partial^{\mu}\phi^k - \frac{1}{2}\mu_{0}^{2}
\phi^k\phi^k) - \frac{\lambda}{4!N}(\sum_{k=1}^{N}\phi^k\phi^k)^2,
\end{equation}
where $\phi$ is an N-dimensional vector and the dimension of spacetime is
3 + 1.  Clearly, the model possesses the O(N) symmetry
with respect to rotation in $\phi$.
This model is of interest not only from the academic viewpoint but also
because for specific N this model is in fact the Higgs sector of Standard Model.
It was studied in many papers [13, 25-31]
(see also \cite{Zinn}). It is well known that this model cannot be considered as a consistent
interacting model due to the problem of triviality. As shown in [26-29] in the leading $\frac{1}{N}$
approximation the effective potential for $\phi$ is double-valued for small and complex for large values
of the scalar field. Furthermore, Re $V(\phi_c) \to -\infty$ as $\phi_c \to \infty$ ($\phi_c = <0|\phi|0>$). One could
expect that the model with a finite cut-off is a sensible theory with effective potential bounded from
below. However, as shown in \cite{NS} although for small values of $\phi_c^2$ the potential of such a model is
real and single-valued, there is the second branch of the effective potential at large values of $\phi_c^2$ on
which the potential is complex and Re $V(\phi_c) \to -\infty$ as $\phi_c \to \infty$. Thus, the O(N) model with finite
cut-off cannot be considered as a completely consisitent theory. Nonetheless, phenomenologically it is a
viable model for small energies.

As follows from Lagrangian (12) every vertex gives an
additional factor $\frac{1}{N}$ in scattering amplitude except, as well known (see, e.g., \cite{Colem}),
the tadpole corrections to the propagator of $\phi$. Therefore, it is
convenient to rearrange perturbation theory so that the tadpole corrections were consistently
taken into account in the lowest order
of perturbation theory. It can be done either by introducing auxiliary field or by using the propagator with
tadpole corrections included.

To calculate the effective action for the elementary field
$\phi$ and the local composite operator $\phi^2$, we first introduce the sources
$J_1 \phi + J_2 \phi^2$. Then we should calculate $W(J_1, J_2)$. After that by performing
the Legendre transformation in $J_1$ and
$J_2$, we get the effective action $\Gamma(\phi_c, \rho)$, where $\phi_c = \frac{\delta W(J_1, J_2)}{\delta J_1} =
<0|\phi|0>$ and $\rho = \frac{\delta W(J_1, J_2)}{\delta J_2} =
<0|\phi^2(x)|0>$. Actually we obtain $\Gamma(\phi_c, \rho)$ by integrating the obvious relation $\frac{\delta \Gamma}
{\delta \rho} = - J_2$, i.e.
\begin{equation}
\Gamma(\phi_c, \rho) = - \int J_2(\rho) d \rho d^4 x + \tilde{\Gamma}(\phi_c),
\end{equation}
where $\tilde{\Gamma}(\phi_c)$ is a constant integration which depends only on $\phi_c$ and can be calculated in the
standard way for $J_2 = 0$ by calculating the effective action only for the field $\phi$.
Of course, the most important part is to get a relation between $J_2$ and $\rho$ or $J_2$ and $M$, where $M$ is
the mass of the field $\phi$. To obtain the effective potential, we set in what follows $\phi_c = const$ and $\rho =
const$.
As in the GN model it is more convenient to use the mass $M$
as an independent variable instead $\rho$
\begin{eqnarray}
\rho = <0|\phi^2(x)|0> &=& \phi^2_c (x) + G(x, x) = \phi^2_c (x) + N \int \frac{d^4 k}{(2\pi)^4}\frac{i}{k^2 - M^2} =
\nonumber\\
& & \phi^2_c (x) + \frac{N}{16\pi^2}(\Lambda^2 - M^2\mbox{ln} (\frac{\Lambda^2}{M^2} + 1)),
\end{eqnarray}
where $G(x, y)$ is the propagator of the field $\phi$ ($G(k) = \frac{i}{k^2 - M^2}$).
According to the method described in the previous section the
sought relation between $J_2$ and $M$ can be obtained from the SD equation in the theory with bare term $J_2 \phi^2$.
Indeed, suppose
we calculated the CJT effective action $\Gamma_{CJT} (\phi_c, G_{CJT})$ for $\phi$ and its propagator $G_{CJT}$
in the theory with that additional bare
term. What is the physical meaning of the CJT effective action at the extremum in $G_{CJT}$?
Obviously, it is minus vacuum energy integrated over spacetime
of the model with the additional bare term and constrained value $<0|\phi|0> = \phi_c$ of the field $\phi$.
Consequently, the corresponding SD equation
$\frac{\delta \Gamma_{CJT} (\phi_c, G_{CJT})}{\delta G_{CJT}} = 0$
gives us the required relation between $J_2$ and $M$ (at extremum the propagator $G_{CJT}$
coincides with the propagator $G$)
\begin{equation}
M^2 = \mu_{0}^2  - 2J_2+ \frac{\lambda}{6N}(\phi_c^2 + \frac{N}{(2\pi)^4} \int \frac{d^4 k}{k^2 + M^2}).
\end{equation}
Thus, we have $\Gamma(\phi_c, M) = \int d^4 x (-\Omega(\phi_c, M))$ (also
$\tilde{\Gamma}(\phi_c) = - \int d^4 x V_{eff}(\phi_c)$), where $\Omega$ is the sought effective potential
\begin{equation}
\Omega(\phi_c, M) = \int J_2(M) \frac{d \rho}{M} dM + V_{eff}(\phi_c).
\end{equation}
By integrating in M, we get the sought action
\begin{eqnarray}
\Omega (\phi_c, M) = - \frac{\lambda}{12}(\phi_c^2 + \frac{N\Lambda^2}{16\pi^2})\frac{M^2}{16\pi^2}\mbox{ln}
(\frac{\Lambda^2}{M^2} + 1) + \nonumber \\
\frac{\lambda N}{24}(\frac{M^2}{16\pi^2})^2\mbox{ln}^2
(\frac{\Lambda^2}{M^2} + 1) - 
\frac{\mu_0^2 N M^2}{32\pi^2}\mbox{ln}
(\frac{\Lambda^2}{M^2} + 1) - \nonumber \\
\frac{N}{64\pi^2}\left[M^4\mbox{ln}(\frac{\Lambda^2}{M^2} + 1) - M^2\Lambda^2 +
\Lambda^4\mbox{ln}(\frac{M^2}{\Lambda^2} + 1) \right] +
V_{eff}(\phi_c).
\end{eqnarray}
It remains to calculate $V_{eff}(\phi_c)$. As we mentioned above $V_{eff}(\phi_c)$
is the effective potential for $\phi$ in the
case $J_2 = 0$. Obviously, it can be calculated also from the CJT eftective action taken at the extremum but in this
case in the theory without additional bare term $J_2 \phi^2$. The CJT effective action
in the O(N) model in the two-loop approximaion is equal to
\begin{eqnarray}
 \Gamma_{CJT}(\phi_c, \tilde{G}) &=& I(\phi_c) +  \frac{i}{2}Tr\mbox{Ln}\tilde{G}^{-1} +\frac{i}{2}TrD^{-1}(\phi_c)
 \tilde{G} \nonumber\\
 &-& \frac{\lambda}{4!N}\int d^4x \tilde{G}_{kk}(x,x)\tilde{G}_{ll}(x,x),
 \end{eqnarray}
 where
 \begin{eqnarray*} 
 I(\phi_c) = \int d^4x (\frac{1}{2}\partial_{\mu}\phi_c\partial^{\mu}\phi_c - 
 \frac{1}{2}\mu_{0}^{2}\phi_c^2- \frac{\lambda}{4!N}(\phi_c^2)^{2}),
 \end{eqnarray*}
 \begin{eqnarray*}
 D^{-1}_{kl} (\phi_c; x, y) = -i \frac{\delta^2I(\phi_c)}{\delta\phi_c^k(x)
\delta\phi_c^l(y)} = -i
(-\partial^{\mu}\partial_{\mu}  - \mu^2_0 - \frac{\lambda}{6N}\phi_c^2
) \delta_{kl},
\end{eqnarray*}
where we keep only terms at the leading order in $\frac{1}{N}$ and $\tilde{G}$ is the propagator of $\phi$
in the theory without the additional bare term $J_2 \phi^2$.
By solving the SD equation
\begin{equation}
\frac{\delta\Gamma_{CJT}(\phi_c, \tilde{G})}{\delta \tilde{G}_{kl}} = 0
\end{equation}
and substituting the found propagator $\tilde{G}_{kl}$ back in (18), we find the effective action
for $\tilde{\Gamma}(\phi_c)$.
To get the effective potential, it suffices to set $\phi_c$ = constant.
The solution of (19) is
$\tilde{G}_{kl} = \frac{i\delta_{kl}}{p^2 - m^2}$, where $m^2$ as a function of $\phi_c$ is given by the equation
\begin{equation}
m^2 = \mu_{0}^2 + \frac{\lambda}{6N}(\phi_c^2 + I),
\end{equation}
where $I = \frac{N}{(2\pi)^4} \int \frac{d^4 k}{k^2 + m^2}$.
Consequently, by substituting $\tilde{G} = \frac{i}{p^2 - m^2}$ in (18), we get
\begin{equation}
V_{eff}(\phi_c) = \frac{\mu_{0}^2}{2}\phi^2_c + \frac{\lambda}{4!N}(\phi^2_c)^2 + 
\frac{N}{2} \int \frac{d^4 k}{(2\pi)^4} \mbox{ln}(1 + \frac{m^2}{k^2}) - 
\frac{\lambda}{4!N} I^2.
\end{equation}

Thus, we have calculated the effective potential for the elementary scalar field $\phi$ and the local composite operator
$\phi^2$ (17) in the O(N) model and showed how the method for calculation of the effective potential for local composite
operators described in the previous section for fermion fields can be generalized to the case of scalar fields. Note that
as one would expect for the effective potential for local composite operators
the effective potential (17) is quadratically divergent. It is finite only at the extremum in $M$ (i.e. when
$J_2 = 0$) and if, of course,
the usual renormalization of mass, coupling constant, and the field $\phi$ is performed.
There is also an important difference between the effective potential (17) and the effective potential for
the composite field $\bar{\psi}\psi$ in the GN model. In the GN model $\bar{\psi}\psi$ is an order parameter and at $g > 0$
in a near-critical region the composite field $\bar{\psi}\psi$ describes the dynamics of light physical particles
(fermion-antifermion bound states). Since $\phi^2$ is not an order parameter in the O(N) model, such an interpretation
of the field $\rho$ and the effective potential (17) is absent in this case. 

To study symmetry breaking in the O(N) model, it is enough to consider the effective potential for the
scalar field $\phi$. Obviously, the effective potential for $\phi$ is minus $\Gamma(\phi_c, M)$
at the extremum in $M$. The extremum of $\Gamma(\phi_c, M)$ in $M$ (or what is the same the extremum of
$\Gamma(\phi_c, \rho)$ in $\rho$) means that $J_2$ = 0. Therefore, as follows from (17) the effective potential for
the field $\phi$ is simply $V_{eff}(\phi_c)$, i.e. it coincides with the effective potential for the
field $\phi$ given by the CJT method at the extremum in $\tilde{G}$. This shows once again a close relation between the
effective potential for local quadratic composite operators and the CJT effective potential. Of course, this is  due to the
fact that one can calculate the generating functional by using the CJT method in the theory with the corresponding bare
terms. Since the effective potential (21) for the field $\phi$ coincides with the effective potential
calculated by Nunes and Schnitzer \cite{NS}, their analysis of the O(N) model in the leading $\frac{1}{N}$
approximation and conclusions remain intact.

\vspace{1cm}

\centerline{IV. Conclusion.}

\vspace{1cm}

Since the effective potential for local composite operators can be calculated by using
the CJT effective potential taken at
extrema in the theory with additional source terms, divergences of the effective potential for local composite
operators are the same as in the CJT effective potential taken at extrema in the theory with the corresponding source
terms. Therefore, we believe that the negative attitude with respect to the effective potential for local composite
operators have to be reconsidered. As follows from our analysis in the GN and the O(N) models
the effective potential for local composite operators is especially useful in
studing symmetry breaking when the corresponding local composite operators are order parameters of theory
(as we saw on the example of the
O(N) model  if they are not order parameters, then the effective potential for composite operator
at least in the leading $\frac{1}{N}$
approximation does not have particular advantages over the usual effective
potential for the elementary scalar field $\phi$ in studing symmetry breaking).

We showed that the auxiliary field method successfully works only in the cases where auxiliary
field is an interpolating field for some physical particles. For example,
in the Gross--Neveu model the effective potential for the composite operator
$\bar{\psi}\psi$ and the effective potential
for auxiliary field describe the same physics at $g > 0$, where the auxiliary field $\sigma$
is an interpolating field for the fermion-antifermion bound state. The situation is different at $g < 0$, where the
auxiliary field $\sigma$ does not describe the dynamics of any physical states of the system
(however, even in this case the auxiliary field can be useful because its propagator describes fermion-antifermion
scattering). On the other hand, the effective potential for the
composite operator $\bar{\psi}\psi$ is a monotonously increasing function for $g < 0$
(a reasonable behavior in the case of repulsion in the fermion-antifermion channel).
The difference in the behavior of the two potentials
is related to the fact that the two potentials
have different physical status. In general, the effective potential for auxiliary field is physically meaningful only at
stationary points. On the other hand, the effective potential for local composite operators is physically  meaningful
at all points where it is the energy density of the model under consideration with the corresponding constrained values
of composite operators. Thus, we consider various effective actions as complementary and useful for the corresponding
problems under consideration.

The author is grateful to Prof. V.A. Miransky for helpful discussions
and valuable remarks and acknowledges useful comments on the text of Prof. A.A. Natale.
The author thanks Profs. Nunes and Schnitzer for bringing his attention to \cite{NS}.
This work was supported in
part by the prize of the President of Ukraine for young scientists for
1998 year and FAPESP grant No. 98/06452-9.


\begin{thebibliography}{99}

\bibitem{Sym} K. Symanzik, Commun.Math.Phys. 16 (1970), 48.
\bibitem{Colema} S. Coleman, in "Laws of Hadronic Matter, 1973 International School of
Subnuclear Physics "Ettore Majorana,'' (A. Zichichi, Ed.), \\
p. 139, New York, 1975.
\bibitem{Banks} T. Banks and S. Raby, Phys.Rev. D14 (1976), 2182.
\bibitem{Ell} U. Ellewanger, Nucl.Phys. B207 (1982), 447.
\bibitem{CJT} J.M. Cornwall, R. Jackiw, E. Tomboulis, Phys.Rev. D10 (1974), 2428.
\bibitem{Haymaker} R.W. Haymaker, Riv.Nuovo Cim. 14, No.8 (1991), 1.
\bibitem{GMK} V.P. Gusynin, V.A. Miransky, and V.A. Kushnir, Phys.
 Rev. D39 (1989), 2355.
\bibitem{mirbook} V. A. Miransky, "Dynamical Symmetry Breaking in
Quantum Field Theories,'' \\
World Scientific, Singapore, 1993.
\bibitem{Mir} V.A. Miransky, Int.J.Mod.Phys. A8 (1993), 135.
\bibitem{MYa} V.A. Miransky and K. Yamawaki, Phys.Rev. D55 (1997), 5051.
\bibitem{GMSh} V.P. Gusynin, V.A. Miransky, and A.V. Shpagin, Phys.Rev. D58 (1998), 085023. 
\bibitem{Gross} D.J. Gross and A. Neveu, Phys.Rev. D10 (1974), 3235.
\bibitem{Cole} S. Coleman, R. Jackiw, and H.D. Politzer, Phys.Rev. D10 (1974), 2491;
\nonumber \\
H.J. Schnitzer, Phys.Rev. D10 (1974), 1800; \nonumber \\
L. Dolan and R. Jackiw, Phys.Rev. D 9 (1974), 3320; \nonumber \\
see also Appendix in [1].
\bibitem{Zinn} J. Zinn-Justin, hep-th/9810198. 
\bibitem{Zam} A.B. Zamolodchikov and A.B. Zamolodchikov, Ann. Phys. (N.Y.) 120 (1979), 253.
\bibitem{Hay} R.W. Haymaker, T. Matsuki, and F. Cooper, Phys.Rev. D35 (1987),
2567.
\bibitem{Hub} J. Hubbard, Phys.Rev.Lett. 3 (1959), 77.
\bibitem{Stra} R.L. Stratonovich, Sov.Phys.-Dokl. 2 (1958), 416.
\bibitem{Haag} R. Haag, Phys.Rev. 112 (1958), 669; \nonumber\\
K. Nishijima, Phys.Rev. 111 (1958), 995; \nonumber\\
W. Zimmermann, Nuovo Cimento 10 (1958), 597.
\bibitem{Pes} M. Peskin, in "Les Houches 1982,'' (J.B. Zuber and R. Stora Eds.), \\
North-Holland, Amsterdam, 1984.
\bibitem{HP} R. Haymaker and J. Perez-Mercader, Phys.Rev. D27 (1983), 1352.
\bibitem{HM} R. Haymaker and T. Matsuki, Phys.Rev. D33 (1986), 1137.
\bibitem{Casa} R. Casalbuoni, S. de Curtis, D. Dominici, and R. Gatto, Phys.Lett. B140 (1984), 354. 
\bibitem{Inoue} M. Inoue, H. Katata, T. Muta, and K. Shimizu, Prog.Theor.Phys. 79 (1988), 519.
\bibitem{Root} R.G. Root, Phys.Rev. D10 (1974), 3322.
\bibitem{Kob} M. Kobayashi and T. Kugo, Prog.Theor.Phys. 54 (1975), 1537.
\bibitem{Abb} L.F. Abbott, J.S. Kang, and H.J. Schnitzer, Phys.Rev. D13 (1976),
2212.
\bibitem{Lin} A.D. Linde, Nucl.Phys. B125 (1977), 369.
\bibitem{Bar} W.A. Bardeen and M. Moshe, Phys.Rev. D28 (1983), 1372.
\bibitem{Ste} P.M. Stevenson, Phys.Rev. D32 (1985), 1389.
\bibitem{NS} J.P. Nunes and H.J. Schnitzer, Int.J.Mod.Phys. A10 (1995), 719.
\bibitem{Colem} S. Coleman, "Aspects of Symmetry,'' \\
Cambridge University Press,  1985.
\end{thebibliography}
\end{document}